# A biomimetic feedback loop for sustaining self-lubrication and wear resistance

Fuyan Kang[1,3,6], Shilin Deng[2,6], Panpan Li[1,6], Rui Zhao[4], Xiaohong Liu[1,3], Hongxuan Li[1,3*], Huidi Zhou[1], Jianmin Chen[1], Wengen Ouyang[2,5*] & Li Ji[1,3*]

1 State Key Laboratory of Solid Lubrication, Lanzhou Institute of Chemical Physics, Chinese Academy of Sciences, Lanzhou 730000, P. R. China.

2 Department of Engineering Mechanics, School of Civil Engineering, Wuhan University, Wuhan, Hubei, 430072, P. R. China.

3 Center of Materials Science and Optoelectronics Engineering, University of Chinese Academy of Sciences, Beijing 100049, P. R. China.

4 School of Mechanical and Electrical Engineering, Xinyu University, Xinyu 338004, P. R. China.

5 State Key Laboratory of Water Resources Engineering and Management, Wuhan University, Wuhan, Hubei, 430072, P. R. China.

6 These authors contributed equally: Fuyan Kang, Shilin Deng, and Panpan Li.

E-mail: lihx@licp.cas.cn; w.g.ouyang@whu.edu.cn; jili@licp.cas.cn

## Abstract

Intelligent materials that self-sense and self-regulate are an emerging frontier in sustainable technology. Here we introduce Cu(Au)/C nanocomposite films that act as bioinspired self-adjusting lubricants. In these films, frictional heating triggers melting and migration of soft metal nanoparticles (NPs) such as Cu or Au along nano-pores to the friction interface, where the metal catalyzes the *in-situ* formation of ordered carbon nano-structures. Real-time monitoring of friction coefficient ($\mu$), electrical resistance (R), and metal release confirms an autonomous cycle: high $\mu$ generates heat, melting the metal NPs; the migrating metal then lowers $\mu$ by creating low-friction nanostructures, which reduces heat and arrests further migration until friction rises again. This self-limiting feedback enables stable ultra-low friction ($\mu$~0.04) and an exceptional wear life (>40 km) even in high vacuum. By utilizing friction-derived heat as an intrinsic activation signal, our system establishes a general paradigm for intelligent, self-regulating materials with applications extending beyond tribology.

# Introduction

Intelligent materials represent a frontier in sustainable technology, with the global market projected at 72.36 billion dollars in 2023, and it is expected to reach 133.1 billion dollars by 2030[1]. These materials offer exceptional energy-saving potential and performance advantages across fields such as advanced manufacturing[2-5], biomedical engineering[6-8], and aerospace[9,10]. In tribology, intelligent lubricants are particularly critical, as they adaptively reduce friction and wear under varying conditions–addressing a key limitation of conventional lubricants[11-15].

Current design strategies–such as biomimetic structures[16-20], microencapsulation[21-23], stimulus-responsive molecules[24-26], and porous structures[27,28]–have led to significant advances. Yet, most systems still operate through passive, mechanical responses rather than true, bio-like intelligence. For instance, some materials aim to use the stability of interface structures to achieve adaptability[29], or some self-healing lubricants require external force or light to release healing agents, often resulting in over-release[22] and inefficient performance[27,30]. This results in the lubricating material being able to only achieve one-time use in the aforementioned situation, without feedback loop. Realizing lubricants with bioinspired, self-sensing, and self-regulating capabilities remains an unmet challenge[31]–one whose solution could extend material lifetimes and enable operation in extreme environments such as vacuum.

Here, we report Cu(Au)/C nanocomposite films that exhibit truly bioinspired intelligent lubrication. Drawing on two known phenomena–friction-induced metal migration[32,33] and metal-catalyzed carbon ordering[33,34]–we designed a material capable of self-adjustment based on real-time tribological conditions. Through *in situ* monitoring $\mu$, R, and metal release, combined with experimental and simulation analysis, we elucidated a feedback mechanism wherein frictional heat mediates metal NP phase transitions and interfacial migration. This system achieves unprecedented wear life and ultralow friction in vacuum (>40 km, $\mu$~0.04), resolving the long-standing failure problem of carbon materials under such conditions. Our work offers a general design strategy for intelligent materials with broad interdisciplinary impact.

# Results

**Construction of bioinspired intelligent Cu(Au)/C lubrication films**

Doped soft metals Cu or/and Au tend to migrate toward friction interfaces, achieving metal supplement thereat[35-38], and catalyze amorphous carbon (a-C) to form special ordered nano-structures, which reducing friction and wear[34,39,40]. Herein, we constructed Cu(Au)/C films by doping Cu or Au into (a-C) matrices to achieve intelligent lubrication, with the expectation of utilizing the two phenomena that migrating metals cognitively provide catalytic units and form low-friction nanostructures (Fig. 1a), like organisms make decisions through neural networks.

The exciting results show that constructed Cu(Au)/C films exhibit highly sensitive bioinspired lubrication behavior detected by real-time multi-parameter monitoring (Fig. 1b and Supplementary Fig. 1), where there are remarkable correlations among $\mu$, R, and metal migration. Initially, a high $\mu$ appears due to strong C-C interactions at the friction interface[41]. Concurrently, the high amount of metal release and R indicate metal migration, similar to that observed in numerous experiments (Supplementary Fig. 2-6). When metal migrates to the friction interface and forms lubricating structures, friction decreases, corresponding the low metal release and R. Combining with corresponding changes in $\mu$, R, and metal release, it is inferred that Cu(Au)/C films exhibit rapid self-adjusting ability when poor lubrication occurs at the friction interface. For example, after the low-friction interface formed by metal migration is artificially disrupted, the metal NPs replenish toward the friction interface which returns to a stable state with the reconstruction of the low-friction structures thereat (Supplementary Fig. 7). This is supported by the correlated changes of the $\mu$, R, and metal release, which confirms the intelligent lubrication behavior. This provides experimental evidence *in situ* that like living organisms, the Cu(Au)/C films can quickly sense the friction state and adjust the metal migration behavior to repair the poor lubricity at the friction interface. Namely, our constructed Cu(Au)/C films exhibit self-sensing, self-adjusting, and self-repairing features based on the metal migration and interfacial catalysis behaviors.

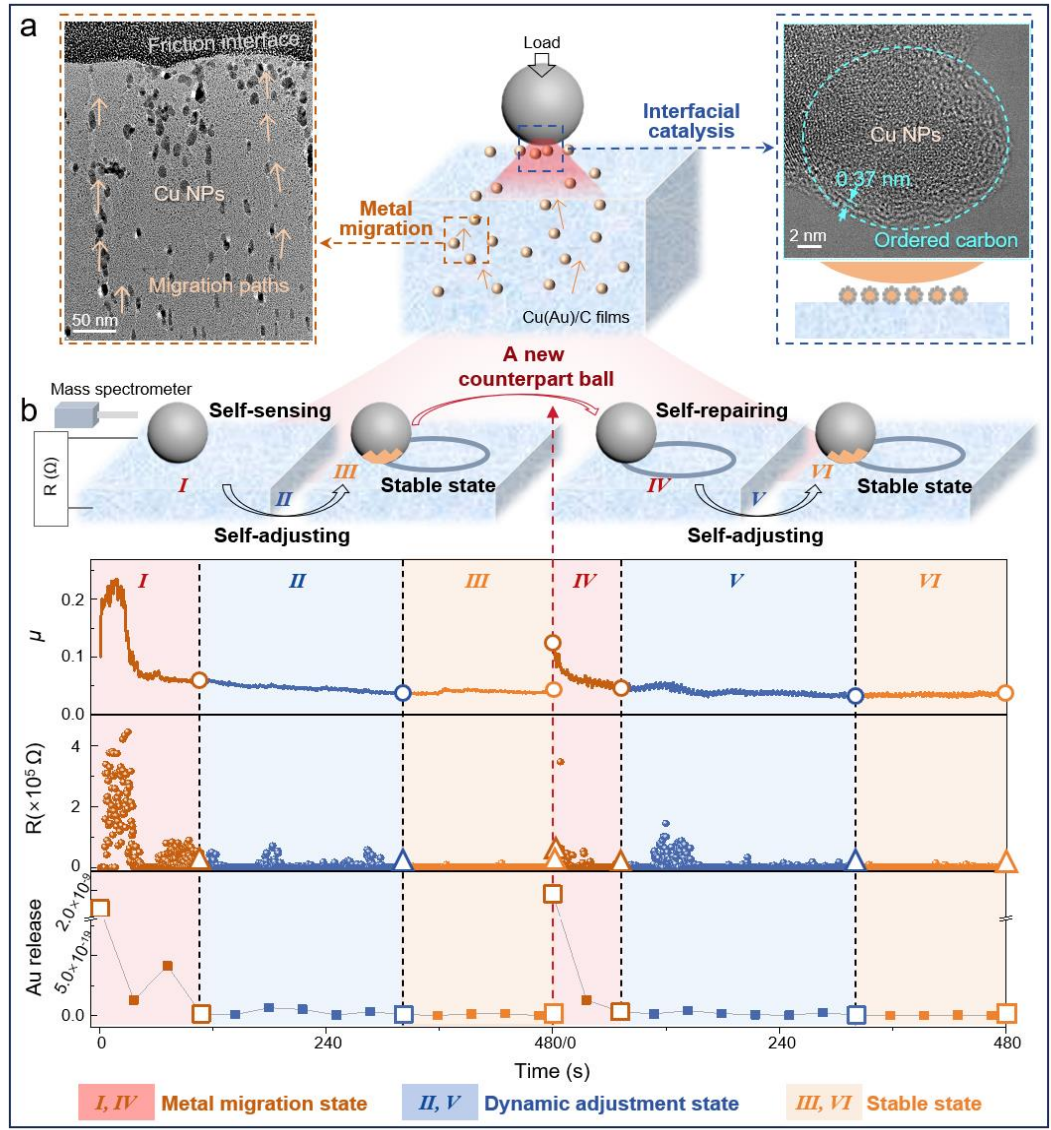

**Fig. 1. Intelligent lubrication of Cu (Au)/C films and real-time observation:** (a) Schematic of Cu(Au)/C intelligent lubrication films design and (b) Schematic of Cu(Au)/C intelligent lubrication films design (The values of Au release refer to the amount of metal released per unit distance and per unit time, expressed in the unit of Amu·mm$^{-1}$·s$^{-1}$).

**Frictional heat-induced solid-liquid phase transition and migration mechanism of metal NPs**

To reveal the bioinspired feedback mechanisms of the Cu(Au)/C films, we conducted in-depth research to comprehend the mechanisms of metal migration and catalysis, respectively. Experiments indicate that the behavior of metal migration is sensitive to heat (Supplementary Fig. 8 and 9). Further, the migration process of Cu NPs in Cu/C films during heating is captured by *in situ* transmission electron microscopy (TEM; Fig. 2a, b, d; Supplementary Fig. 10, 11, and Supplementary Video 1). With increase in temperature, the Cu atoms agglomerate and Cu NPs grow due to energy minimization[42,43]. Notably, spherical Cu NPs appear on the surface of Cu/C films heated at 300 °C as marked by the orange boxes (Fig. 2b). This is attributed to the migration of Cu NPs onto the nearest surface, which is consistent with the phenomenon of Cu precipitation on the Cu/C film surface after macro heating (Supplementary Fig. 9). Unexpectedly, when the temperature rises to 400 °C, the newly grown Cu NPs gradually "disappeared" (Supplementary Fig. 12). Considering the decrease of Cu NPs' melting temperature due to the nano effect[44], we measured the melting temperature of Cu NPs in Cu/C films using differential scanning calorimetry-thermogravimetric analysis (DSC-TG). As shown in Fig. 2c, different reactions occur in the Cu/C film and the C film based on occurring peaks and the mass loss of films. In combination with the *in situ* TEM observation of the disordered Cu NP during holding at 300 °C (Fig. 2d) and further X-ray photoelectron spectroscopy (XPS) of Cu/C film powders after DSC-TG test (Supplementary Fig. 13), it is inferred that the endothermic peak at 270°C is attributed to the melting of Cu NPs in the Cu/C film. Besides, the exothermic peak in the DSC curves of Cu/C film may be ascribed to the crystallization of molten Cu NPs of Cu/C powders upon volatilization, accompanied by significant mass loss. Fig. 2d illustrates the real-time migration process of the liquid Cu NP toward the surface. Relevant inverse fast Fourier transform (IFFT) images of the Cu (200) and Cu (111) planes of the Cu/C film holding at 300 °C demonstrate that the Cu NP undergoes amorphization, referring to its melting. The IFFT images of C (a-C) indicate that the contrast of the C film surface change significantly over time, which refer to the precipitation of Cu NPs onto the film surface. Throughout the process, while amorphous Cu precipitates toward the surface of Cu/C film and gradually crystallizes because of the lower temperature thereon. Ultimately, the Cu NPs with complete crystal structures precipitate on Cu/C film surface.

To elucidate the experimental observations, we conducted molecular dynamics (MD) simulations. During the film deposition, intrinsic defects inevitably formed (Supplementary Fig. 9a), providing migration channels for metal atoms[36]. Since nano-pores in the vary in shape and size, we introduced a conical hole in the a-C matrix to model a typical nano-pore (Fig. 2e). When solid Cu melts into liquid Cu under the applied heat flux, the Cu droplet deforms due to their interaction with surrounding carbon dangling bonds. These droplets are drawn toward the film surface through two complementary mechanisms: attraction by unsaturated carbon bonds and a Laplace pressure gradient arising from the conical pore geometry[45]. Driven by this chemical potential difference, Cu droplets migrate rapidly across long distances through the nano-pores (Fig. 2e, Supplementary Fig. 14, and Supplementary Video 2), enabling efficient supply of metal NPs at the friction interface. This migration mechanism is corroborated by cross-sectional TEM imaging of the wear track, where elliptical Cu droplets are clearly observed (Fig. 2f).

The temperature gradient across the friction interface further modulates this process (Supplementary Fig. 15), giving rise to stepwise migration. Initially dispersed small NPs melt and migrate toward the surface while concurrently coalescing into larger aggregates, leading to a characteristic size gradient. Although sufficiently high frictional heat can melt large aggregates, the narrow pores permit only smaller droplets to pass through. Consequently, larger NPs solidify and remain embedded as metal reservoirs, preserving the film's mechanical stability (Supplementary Fig. 16) despite surface accumulation of metal NPs. These findings establish the phase transition kinetics and transport mechanisms governing metal migration in solid matrices, providing fundamental insights for designing self-regulating metal-carbon composites with tailored surface architectures under thermal and tribological stimulation.

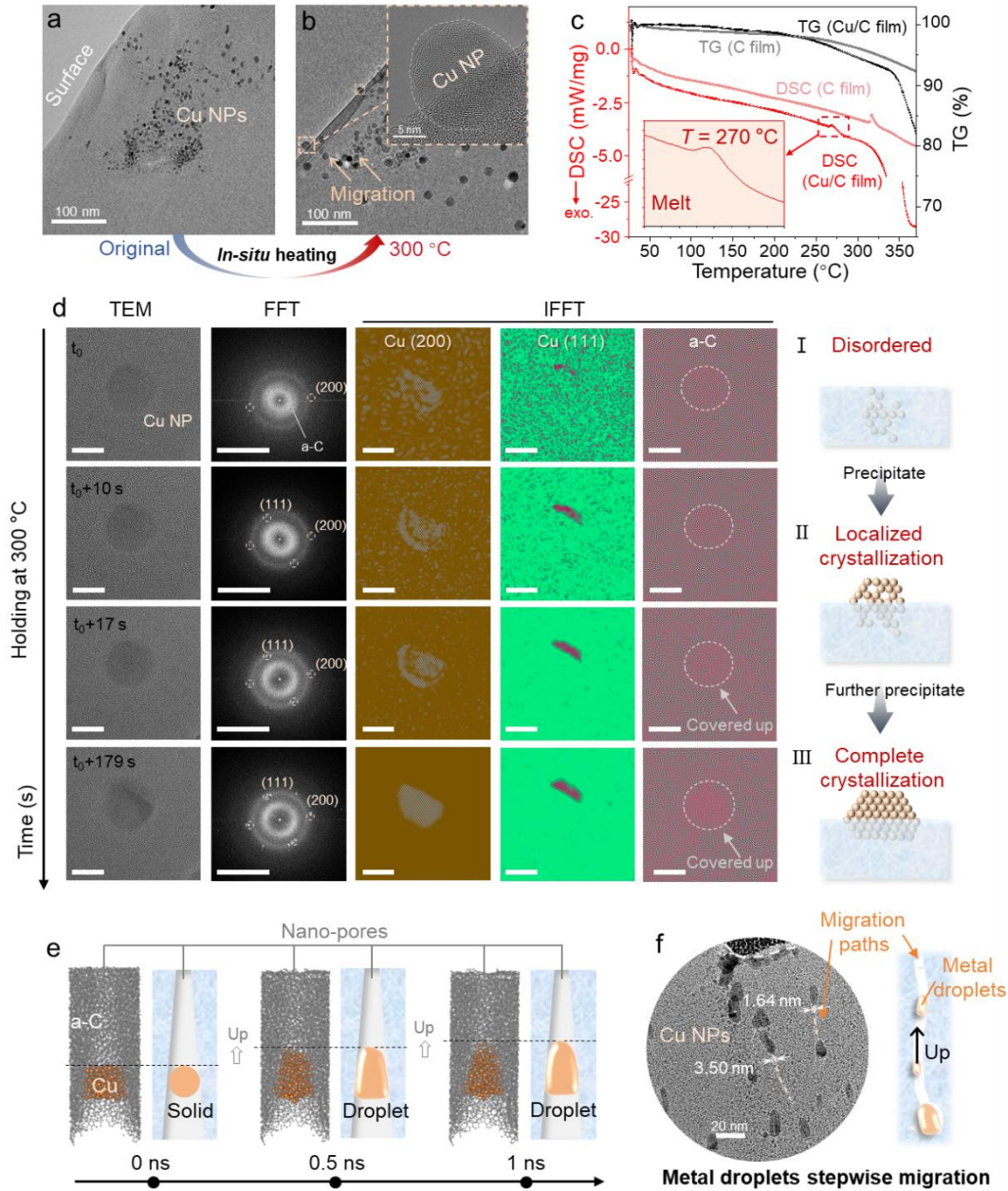

**Fig. 2. Migration mechanism of heat-induced solid-liquid phase transition of Cu NPs through *in situ* TEM observations and MD simulations**: (a-b) *In situ* HRTEM observation of the migration of Cu NPs onto the surface, (c) DSC and TG Curves of C and Cu/C powders to show that the melt temperature of Cu NPs in Cu/C films is about 270 °C, (d) The process of molten Cu NPs precipitating to the surface at 300 °C (Scale bars in the TEM and IFFT images are 10 nm, while scale bars in FFT images is 10 1/nm ), (e) MD simulations of the solid-liquid phase transition and migration of Cu NPs towards the surface along the nano-pores (intrinsic defects of the films), and (f) Heat-induced molten Cu NPs gradually migration during friction (The left image shows a cross-sectional TEM image of the wear track of Cu/C film).

**Metal catalysis forms low-friction ordered nano-structures**

To investigate the metal-catalyzed mechanism, we conducted *in situ* experiments, simulations and calculations. As shown in Fig. 3a-d and Supplementary Video 3, the initially uniformly dispersed small-sized Cu NPs (2~3 nm) inevitably agglomerate into a larger-sized Cu NP (15~20 nm) during heating due to energy minimization. Surprisingly, at 300 °C, the edges of the large Cu NP are wrapped with multiple layers of ordered carbon. Moreover, ordered carbon structures occur around the Cu NPs in the wear debris during the low $\mu$ stage (Fig. 1a). These results demonstrate the transition from $sp^3$C to $sp^2$C under Cu catalysis, which is prone to form carbon-encapsulated metal nanostructures[46]. Ordered carbon-encapsulated metal nanostructures can shield carbon dangling bond interactions[34,40,47], reduce the contact area[48], and utilize configuration characteristics for rolling lubrication at the friction interface[49,50], thereby reducing friction and wear. MD simulations also indicate the migrating metal NPs will form this low-friction ordered carbon-encapsulated Cu nano-structures after the Cu NP moves to the surface during friction (Fig. 3e). Additionally, sufficient Cu is required to provide catalytic units at the friction interface to form these lubricating structures, even though the amount required is minimal (Supplementary Fig. 17). Once enough metal accumulates after the run-in period, low-friction nano-structures are prone to forming at the friction interface and reach a stable lubricating state with increased structural stability (Supplementary Fig. 18).

The catalytic mechanism of Cu NPs is further explored through MD simulations and Density Functional Theory (DFT) calculations. When carbon atoms progressively transformed from tetrahedral $sp^3$ (*I* in Fig. 3f) to graphitic-like $sp^2$ (*IV* in Fig. 3f) coordination with lower energy during the breaking and recombination of C-C bonds (*II* and *III* in Fig. 3f), due to the Cu catalysis (Supplementary Video 4), there was a significant charge rearrangement occurred in the carbon structure before and after the C-C bond breakage (Fig. 3i, j). These results indicate the presence of unique electronic effects at the Cu-C interface that promote the transformation of carbon structures. To quantify the Cu-induced electronic effect, we computed the crystal orbital Hamilton population (COHP) of the target C-C bond with and without Cu NPs. The presence of Cu introduces an antibonding feature near the Fermi level of the C-C bond, indicative of Cu-C anti-bond formation[51] (Fig. 3k and Supplementary Fig. 19 and 20). This effect stems from the lower work function of Cu, which promotes electron donation at the Cu-C interface (Supplementary Fig. 21). The corresponding reduction in the -ICOHP value of the C-C bond confirms that the Cu-C anti-bond weakens the C-C interaction (Fig. 3k) and loosens the carbon network[51]. This weakening facilitates C-C bond cleavage during thermal vibration, ultimately leading to the formation of thermodynamically more stable $sp^2$-hybridized carbon structures on the surface of Cu NPs.

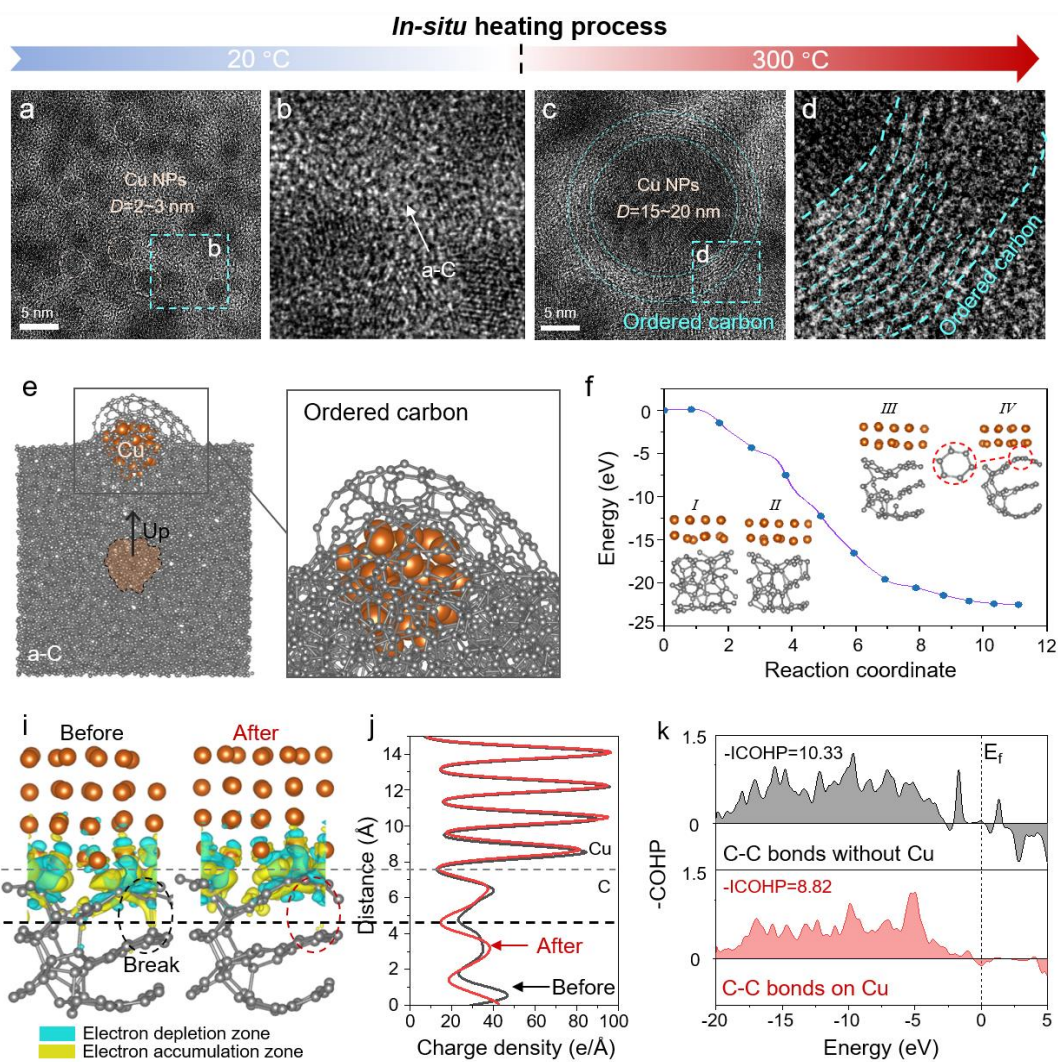

**Fig. 3. Cu catalysis forms ordered nano-structures and catalytic mechanisms:** (a-d) *In situ* TEM images of Cu/C films during heating to show the transform of a-C to ordered carbon, (e) MD simulations of formation of low-friction nano-structures under Cu catalysis after Cu moving toward the surface, (f) Energy changes in the Cu-catalyzed transformation from a-C to ordered carbon (*I*. a-C; *II, III*. $sp^3$C breakage and $sp^2$C recombination; *IV*. Ordered carbon), (i) Before and after the C-C bond breaking during Cu catalysis, (j) Charge density distribution before and after C-C bond breakage, and (k) COHP of the target C-C bond with and without Cu NPs (The lower the absolute value of ICOHP,

the weaker the of the C-C bond strength).

**Feedback mechanisms of the bioinspired intelligent lubricant**

Base on the abovement, the mechanism cyclic feedback interaction among the friction state, frictional heat, and metal migration enables the realization of bioinspired intelligent lubrication (Fig. 4a) is proposed. Initially, high interfacial friction arises due to strong interactions among carbon dangling bonds, generating high frictional heat[52]. Metal migration can be precisely regulated in response to frictional heat owing to the size-dependent melting temperature of metal NPs[53]. Sufficient frictional heat melts metal NPs within Cu(Au)/C films, triggering to their rapid migration toward the friction interface, while insufficient frictional heat prevent this phase transition and migration. The migrating metal NPs can further form low-friction nano-structures under catalysis thereat, thereby reducing friction and frictional heat. When low-friction nano-structures is worn away or damaged by external factors, the friction interface reverts to a high-friction and high-friction-heat state, initiating a new cycle of friction to exhibit a self-adaptive lubrication state. This allows to achieve more advanced intelligent lubrication behavior including self-sensing, self-adjusting, and self-repairing. It is noteworthy that this design utilizes friction byproducts to provide energy without over-consumption, which holds significant implications for energy saving.

Surprisingly, the Cu/C film show robust lubrication with an ultra-long wear life of 40 km in vacuum ($\mu$~0.04), 147 times and 38 times longer than that of hydrogen-free C films and hydrogen-containing C films, respectively (Fig. 4b). This significantly extended lifespan breaks the curse of hydrogen-free C films' poor lubrication in vacuum, without the effects of hydrogen tribo-emission[54], which provides new avenue for avoiding rapid failure of carbon materials in vacuum.

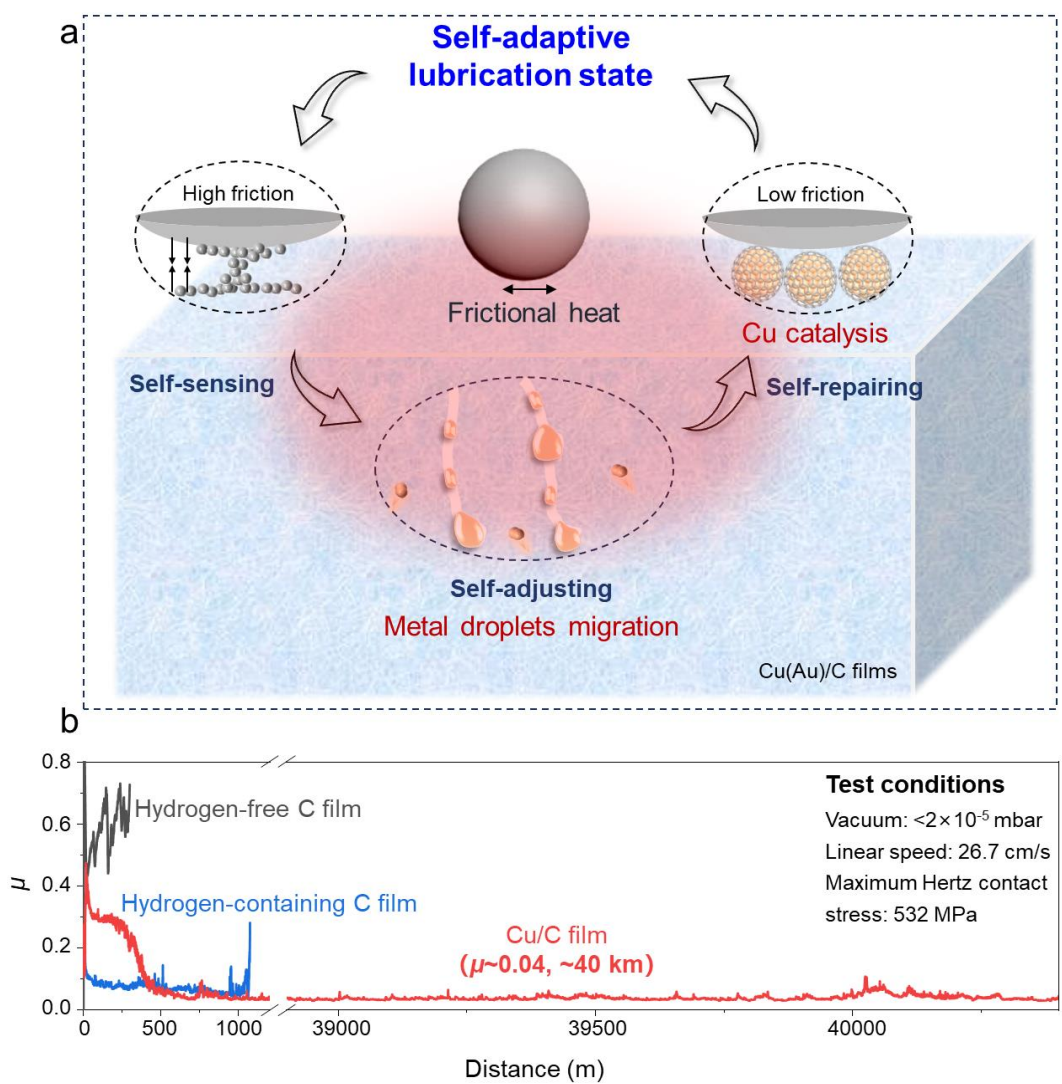

**Fig. 4. Feedback mechanism of intelligent lubrication of Cu(Au)/C films and the excellent lubricating performance in vacuum:** (a) Schematic diagram of the feedback mechanism of the intelligent lubrication system powered by friction byproducts, and (b) The ultra-long wear life of the Cu/C film.

## Discussion

In summary, we have developed Cu(Au)/C nanocomposite films as a new class of intelligent lubricants that exhibit bioinspired autonomous regulation. Real-time monitoring of $\mu$, R and metal release uncovers responsive correlations that reveal self-sensing, self-adjusting and self-repairing capabilities. In situ experiments combined with MD simulations identify the fundamental feedback mechanism: frictional heating induces solid–liquid phase transitions in metal NPs, which migrate to the friction interface and catalyze the formation of low-friction nanostructures. This self-limiting cycle enables outstanding tribological performance, sustaining an ultralow $\mu$ (~0.04) and an ultra-long wear life of over 40 km in vacuum, thereby overcoming the long-standing problem of rapid failure in carbon-based lubricants under such conditions. By converting frictional energy into a driver for autonomous surface regulation, this work establishes a bioinspired design paradigm for intelligent materials with broad potential in tribology, sensing, corrosion protection and advanced manufacturing.

## Methods

### Film preparation

Cu(Au)/C films were prepared by a magnetron sputtering system (PlasMag CF-800, Teer, U.K.) through co-sputtering graphite and metal targets. The films were deposited on Si wafers, 304 stainless steel, and M2 stainless steel. Before film deposition, the substrates were ultrasonically cleaned with anhydrous ethanol and acetone, respectively, and dried with dry $N_2$. The deposition process and information about films are shown in Supplementary Fig. 22 and the literature[32,33].

### Friction and wear tests

Friction and wear tests were conducted with a ball-on-disk tribometer (HVTRB, CSM, Anton Paar, Switzerland) in vacuum (pressure < $2\times10^{-5}$ mbar) with rotating mode at a linear speed of 26.7 cm/s, a maximum Hertz contact stress of 532 MPa. The GCr15 steel balls ($\varphi$6 mm, AISI52100, roughness ≈ 20 nm) were used as the counterpart balls. A mass spectrometer (PrismaPro, Pfeiffer, Germany) and a multimeter (8845A/8846A, Fluke, USA) were connected outside the chamber of the tribometer to detect R and the metal release during friction. The reciprocating module of the tribometer was used for friction testing at an amplitude of 5 mm and a linear velocity of 4.71 cm/s, with which the GCr15 counterpart balls as described above were used. The morphology of wear scars and wear tracks were observed by optical microscopy (STM6, OLYMPUS, Japan) and Field emission scanning electron microscope (FESEM, SU8020, HITACHI, Japan). The element components of wear scars were identified through equipped element distribution scanning (EDS, EDAX, America). The depth of wear tracks was measured via the three-dimensional white light interferometry (UP-Lambda, Rtec-Instruments, America) and further calculated the wear rates.

### Vacuum annealing tests

Annealing experiments were performed by using the heating chamber of the ball-on-disk tribometer

(HVTRB, Anton Paar, Switzerland) in vacuum (pressure $< 2 \times 10^{-5}$ mbar), The sample were held at 100 °C, 200 °C, and 300 °C for 3 hours each to study the metal migration behavior.

**Mechanical performance tests**

The mechanical properties of Cu(Au)/C films were tested by nanoindentation (NHT2, CSM, Anton Paar, Switzerland) and CSM scratch meter (RSTNHT2, Anton Paar, Switzerland). For details, see Supplementary Fig. 23 and the literature[32,33].

**Characterization:**

The original structure of Cu(Au)/C films are shown in Supplementary Fig. 24 and the literature[32,33]. The crystal structure information was obtained by XRD (Empyrean, Maiven Panalytical, Netherland). The carbon structure information was tested by Raman spectroscopy (Renishaw, UK, wavelength = 532 nm). The dynamic structure evolution of a-C during friction was on-line investigated through *in situ* Raman spectroscopy (wavelength = 532 nm). A three-dimensional white light interferometry (MicroXAM-3D, ADE, America) was used to measure the film thickness. The Cu content in Cu/C film was detected via EDS (EDS, Oxford, Britain). The FESEM (SU8020, HITACHI, Japan) was used to observe the morphology of films before and after annealing at different temperatures. The elemental composition of the surface after annealing at 300°C were tested by point scanning using the equipped EDS (EDS, EDAX, America). Details of the static pressure test are shown in Supplementary Fig. 6. Original structures of the Cu/C film and the contact area before and after static pressing were analyzed by *ex-situ* XPS (ESCALAB 250Xi, Thermo Fisher, America), and elemental analysis of powders of the Cu/C film before and after DSC-TG testing was performed. The elemental analysis of the Cu/C film was further characterized by *in situ* XPS (VersaProbe 4, PHI, Japan) at different temperature during heating. The migration behavior of Cu NPs in a-C matrices was observed by *in situ* TEM (JEM-ARM300F2, JEOL, Japan) at 25 frames per second, with an electron beam voltage of 300 eV. Equipped with an electric heating unit for heating the sample with a temperature error of no more than ± 10 °C. The samples were cut from the Cu/C (hydrogen-free) film and Cu/C (hydrogen-containing) film respectively using a focused ion beam (FIB, Helios Nanolab 600i, Thermo Scientific, America). The Cs-corrected TEM (Spectra 300, Thermo Fisher Scientific, America) was applied to observe the morphology of wear debris of the Cu/C film. The cross-sectional morphology of the characteristic wear tracks was studied though TEM (TECNAI G2 F20, FEI, America), which was prepared by the FIB (Helias 5UX, Thermoscientific, America) method. The melting temperature of Cu NPs in Cu/C film was obtained by DSC-TG tests (STA 449 F3 Jupiter, Netzsch, Germany). Ar flow was introduced for a period of time prior to the temperature increase, followed by a temperature increase at 10 °C/min, a holding time of 5 min and then temperature decreases at the same rate.

**Development of the Neuroevolution potential (NEP) model:**

The NEP model[55,56] writes the site energy of a central atom $i$ as:

$$U_i\left(\{q_\nu^i\}_{\nu=1}^{N_{des}}\right) = \sum_{\mu=1}^{N_{neu}} w_\mu^{(1)} \tanh\left(\sum_{\nu=1}^{N_{des}} w_{\mu\nu}^{(0)} q_\nu^i - b_\mu^{(0)}\right) - b^{(1)} \quad (S1)$$

where $\tanh(*)$ denotes the activation function of the hidden layer, and $N_{neu}$ represents the number of neurons. The parameters $w^{(0)}$, $w^{(1)}$, $b^{(0)}$, and $b^{(1)}$ are the weight and bias parameters to be optimized during the training process. The descriptor $q_\nu^i$ consists of radial and angular components. To develop a NEP model capable of accurately describing Cu-C interactions, we reused the training data set constructed by Wang et al[57] and expanded it by creating a binary Cu-C dataset. To construct this dataset, several representative Cu-C structures with varying compositional ratios were selected as initial configurations for ab initio molecular dynamics (AIMD) simulations. These simulations were performed under the NPT ensemble using a Langevin thermostat at 300 K, 450 K, 600 K and 800 K, with a simulation time of 10 ps for each temperature. Cu-C surface models were also constructed and simulated under the NVT ensemble using a Nosé-Hoover thermostat, with the temperature increasing from 300 K to 800 K over 10 ps. After initial datasets are established, DFT single-point energy calculations were performed using Vienna Ab-initio Simulation Package (VASP)[58,59], employing the Projected Augmented Wave (PAW) method and the Perdew-Burke-Ernzerhof (PBE) functional[60]. The calculations used a plane-wave energy cutoff of 550 eV, a $k$-point spacing of 0.2/Å, and an energy convergence threshold of $10^{-6}$ eV. The final DFT reference dataset comprised 1,644 configurations, which were randomly divided into a training set of 1,480 configurations and a validation set of 164 configurations. The training results of the NEP model are presented in Supplementary Fig. 25 and the hyperparameters for training are listed in Supplementary Table 1.

**Model and set-up for MD simulations**

Our MD simulations comprised three parts: (i) migration of metal NPs within a conical pore in a a-C

model, (ii) guided metal NP motion to probe catalytic behavior, and (iii) AIMD simulations to explore Cu-C interfacial interactions. For metal migration, a circular conical frustum (base diameter ≈ 3 nm; top diameter ≈ 1 nm) was carved into the a-C model (model size ≈ 3 × 3 × 7 nm³; Supplementary Fig. 26) with a 227-atom Cu NP placed near at the pore entrance. The initial structure was energy-minimized with FIRE method (force < $10^{-6}$ eV/Å) and further equilibrated at 800 K for 1 ns while fixing the bottom 1 nm in the NVT ensemble (Nosé-Hoover thermostat, 1 fs timestep). For metal migration behavior, simulations were performed in Graphics Processing Units Molecular Dynamics (GPUMD) package[55]. To overcome timescale limitations, the metal NP was driven at 5 m/s along the pore axis in Large-scale Atomic/Molecular Massively Parallel Simulator (LAMMPS)[61] to study the catalysis effect of migrating metal NPs. The simulation utilized a 1 fs timestep, 300 K Nosé-Hoover thermostat. Interfacial interactions during the Cu-catalyzed ordered carbon were further examined by AIMD in VASP (PBE functional, 500 eV cutoff, $10^{-4}$ eV energy convergence). NVT simulations were run at 800 K with a 3 fs timestep for 30 ps. The initial model comprised an amorphous $sp^3$-rich carbon substrate (66 atoms) with a 32-atom Cu NP on top in a 0.7 × 0.7 nm² cell with 3 nm vacuum spacing along the $z$-axis.

**Model and set-up for DFT calculations**

All DFT calculations were performed using VASP[58,59] within the PBE[60] functional with a plane-wave cutoff of 500 eV. Atomic positions were fully relaxed until residual forces were < 0.01 eV/Å and energies converged to $10^{-4}$ eV. Brillouin-zone sampling employed a 7 × 7 × 1 Monkhorst-Pack grid. The charge density distribution was calculated by:

$$\rho_{\text{diff}}(x,y,z) = \rho_{\text{tot}}(x,y,z) - \rho_{\text{Cu}}(x,y,z) - \rho_{\text{C}}(x,y,z) \quad (S2)$$

Where $\rho_{\text{tot}}(x,y,z)$, $\rho_{\text{Cu}}(x,y,z)$, $\rho_{\text{C}}(x,y,z)$ are the charge density of Cu-C, Cu and C.

The climbing image nudged elastic band (CI-NEB) method was used to map the reaction energy landscape, with the optimized pre- and post-catalytic Cu-C structures as the initial and final states, respectively. The reaction path was discretized using 11 images, and the spring constant was set to 5.0 eV/Å². The work function was calculated to characterize the electronic properties of the constructed Cu-C model. The Cu slab was cleaved using a 2 × 2 × 2 surface supercell, with 4 atomic layers and ~20 Å of vacuum along the z-axis. The bottom two layers were fixed to emulate bulk termination, while the top two layers were relaxed to < 0.01 eV Å⁻¹. Brillouin-zone sampling used a 7 × 7 × 1 Monkhorst-Pack grid. Additionality, we employed an a-C slab representative of the substrate. The final slab contained ~66 atoms in the active region, with an in-plane cross-section ~7.5 × 7.5 Å² and ~35 Å of vacuum. All calculations were performed using DFT within the GGA-PBE as implemented in VASP. The work function was extracted as

$$\Phi = E_{\text{vac}} - E_{\text{Fermi}} \tag{S3}$$

where $E_{\text{Fermi}}$ denotes the Fermi level of the system. $E_{\text{vac}}$ denotes the vacuum energy level of the system.

The COHP analysis was performed to quantify bonding and antibonding interactions between specific atom pairs. COHP decomposes the band-structure energy into contributions from selected atom pairs and orbital pairs. Single-point wavefunctions from VASP were post-processed with Local Orbital Basis Suite Towards Electronic-Structure Reconstruction (LOBSTER)[62], which projects plane-wave states onto localized orbital bases for electronic-structure analysis.

**Data availability**

The data supporting the findings of this study are available within the main text and supplementary information files. All data are available from the corresponding author upon request.

## Acknowledgements

The authors are grateful to the Strategic Priority Research Program of the Chinese Academy of Sciences (Grant No. XDB0470202), National Natural Science Foundation of China (Nos. 52405232, 52275222, 12472099 and U2441207), and Fundamental Research Funds for the Central Universities (Nos. 2042025kf0050, 2042025kf0013 and 600460100). Computations were conducted at the Supercomputing Center of Wuhan University and the National Supercomputer TianHe-1(A) Center in Tianjin. The help in the *in situ* TEM observation by C. S. Ma and language polishing by L.G. Yu is greatly appreciated.

## Author contributions

F.Y. Kang, S.L. Deng, and P. P. Li contributed equally to this work. L. Ji, W.G. Ouyang, and H. X. Li conceived the project. F. Y. Kang and P. P. Li prepared the samples and performed the experiments.

W.G. Ouyang designed the MD simulation setup. S.L. Deng and R. Zhao conducted the simulations and related analysis under the supervision of W.G. Ouyang. X. H. Liu, H. D. Zhou, and J. M. Chen helped to analyze the data and discuss the mechanisms. L. Ji, W.G. Ouyang, H. X. Li, and P. P. Li supervised the work. All authors discussed the results and contributed to the final version of the manuscript.

**Competing interests**

The authors declare no competing interests.

# Supplementary Material for

# "A biomimetic feedback loop for sustaining self-lubrication and wear resistance"

Fuyan Kang[1,3,6], Shilin Deng[2,6], Panpan Li[1,6], Rui Zhao[4], Xiaohong Liu[1,3], Hongxuan Li[1,3*],

Huidi Zhou[1], Jianmin Chen[1], Wengen Ouyang[2,5*] & Li Ji[1,3*]

1 State Key Laboratory of Solid Lubrication, Lanzhou Institute of Chemical Physics, Chinese Academy of Sciences, Lanzhou 730000, P. R. China.

2 Department of Engineering Mechanics, School of Civil Engineering, Wuhan University, Wuhan, Hubei, 430072, P. R. China.

3 Center of Materials Science and Optoelectronics Engineering, University of Chinese Academy of Sciences, Beijing 100049, P. R. China.

4 School of Mechanical and Electrical Engineering, Xinyu University, Xinyu 338004, P. R. China.

5 State Key Laboratory of Water Resources Engineering and Management, Wuhan University, Wuhan, Hubei, 430072, P. R. China.

6 These authors contributed equally: Fuyan Kang, Shilin Deng, and Panpan Li.

E-mail: lihx@licp.cas.cn; w.g.ouyang@whu.edu.cn; jili@licp.cas.cn

**This Supplementary Material file includes:**

**Supplementary Note 1: The intelligent lubrication behavior of Cu(Au) nanocomposite films.**

The *in situ* detection device of the intelligent lubrication behavior of the Cu(Au) films is depicted in Supplementary Fig. 1. A mass spectrometer and a multimeter are connected outside the chamber of the tribometer to detect the resistance (R) and the metal release during friction.

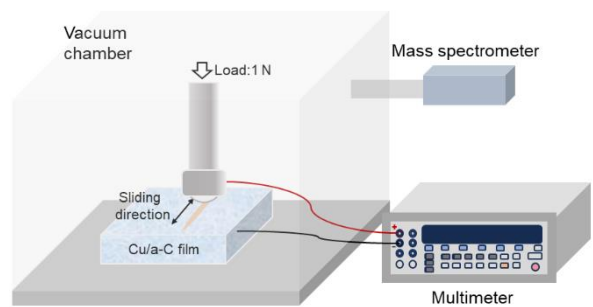

Supplementary Fig. 1| The schematic diagram of the *in situ* detection device.

Supplementary Fig. 2-4 depicts the relationship between friction coefficient ($\mu$) and R at different friction frequencies is shown in. The R exhibits an increasing trend with rising friction frequency. Significant film degradation and rapid failure occur due to over-high high friction frequencies (6.0 Hz and 9.0 Hz). Therefore, the analysis focuses on the tribological behavior at other friction frequencies. Overall, a high $\mu$ promotes metal migration, yielding relatively high R values, while the formation of metal-rich transfer films leads to a low $\mu$ and relatively low R values.

Analysis of wear rates and wear track profiles reveals two distinct friction mechanisms: mechanical wear-induced metal transfer at low friction frequencies and friction-induced metal migration at medium friction frequencies. The high wear rates and ploughing at low friction frequencies of 0.5Hz and 1.5 Hz (0.78 cm/s and 2.36 cm/s) (Supplementary Fig. 2 and 6) are attributed to insufficient frictional heat, which inhibits metal migration. Instead, mechanical wear dominates, causing high wear rates and ploughing in the wear tracks. Despite this, metal-rich transfer films form (Supplementary Fig. 5), resulting in a low $\mu$. At medium speeds of 3.0Hz and 4.5Hz, the wear tracks show no excessive ploughs and lower wear rates, which are attributed to the generation of sufficient heat to cause metal migration, thereby forming a favorable friction interface.

At the employed friction frequency of 3 Hz, the friction process involves the metal migration and the formation of transfer films, both of which affect the overall R value of the system. To simplify the analysis, the friction interface is treated as a single entity for analysis. From Supplementary Fig. 2 and 3, when the friction interface is in a metal-lack state (run-in state), during which the R values are relatively high. While the friction interface is in a metal-rich state (stable state), during which the R values are relatively low. Since mechanical wear-induced metal transfer does not alter the metal distribution significantly, the R value remains relatively stable throughout the friction process.

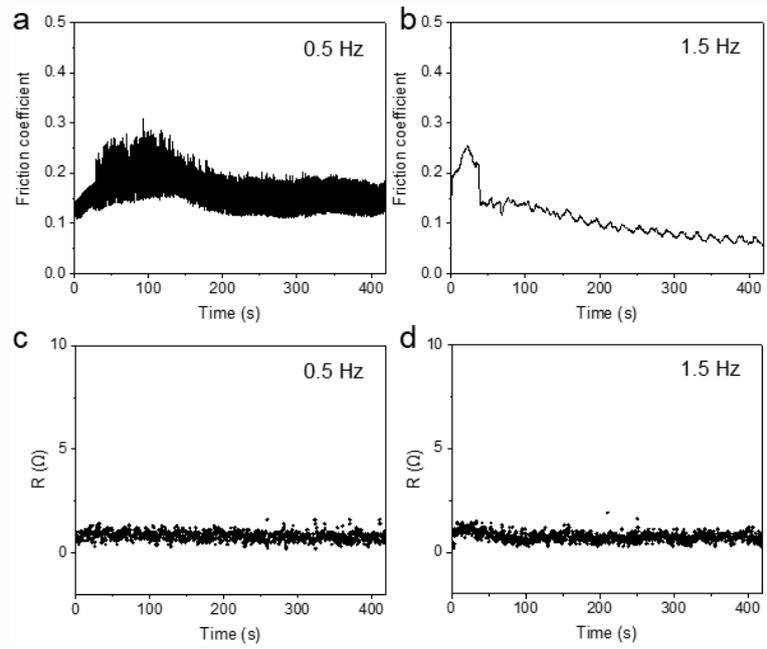

Supplementary Fig. 2| Correlations between $\mu$ and R at low friction frequencies.

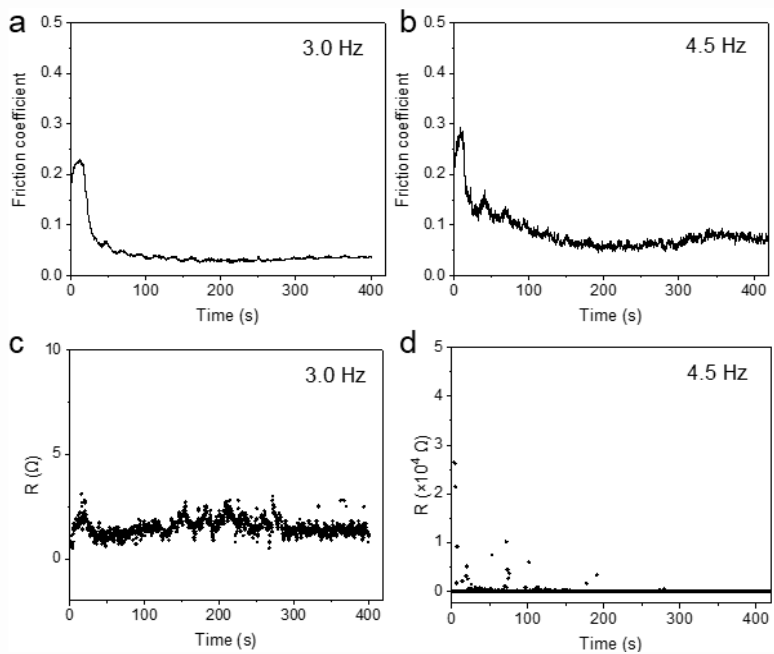

Supplementary Fig. 3| Correlations between the $\mu$ and R at medium friction frequencies.

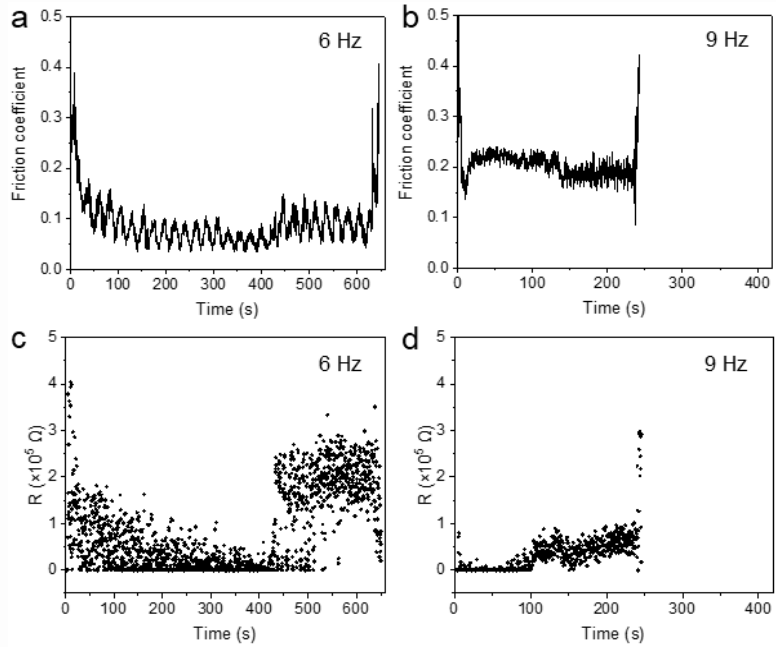

Supplementary Fig. 4| Correlations between *μ* and R at high friction frequencies.

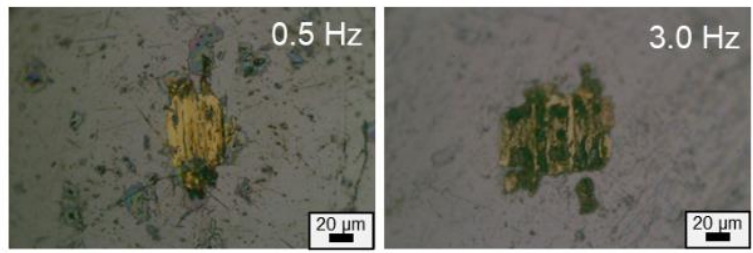

Supplementary Fig. 5| Au-rich transfer films on the counterpart balls under 0.5 Hz and 3 Hz.

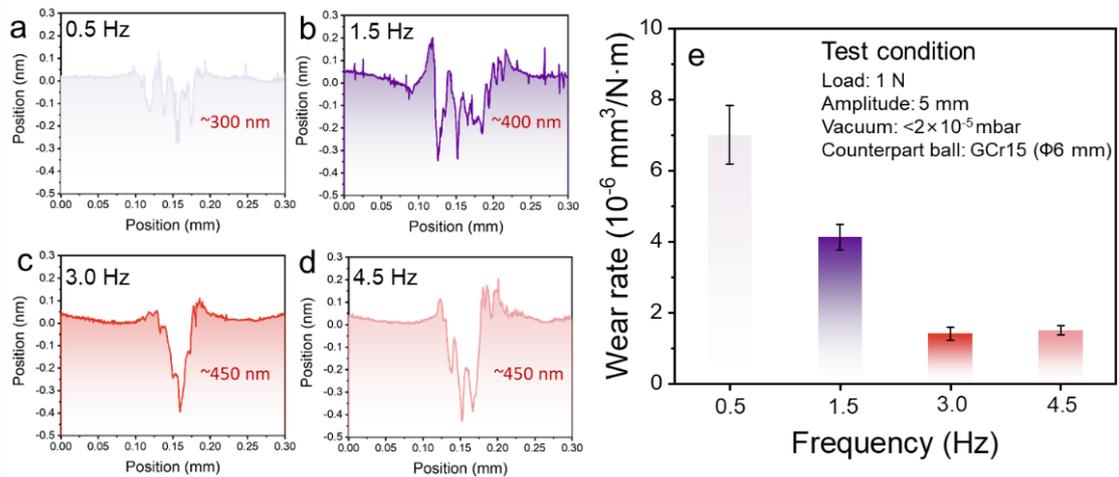

Supplementary Fig. 6| a-d Two-dimensional profiles of the wear tracks under different frequencies. e Wear rates of the Au/C film under different frequencies.

To confirm the self-repairing intelligent lubrication of the Cu(Au)/C films, we artificially disrupted the stable friction interface. After a low-friction, metal-rich transfer films formed on the counterpart ball (Supplementary Fig. 7a), we replaced the counterpart ball with new one. The film responded by rapidly replenishing metal nanoparticles (NPs) to the friction interface and re-forming metal-rich transfer films (Supplementary Fig. 7b), thereby re-establishing a stable lubrication state.

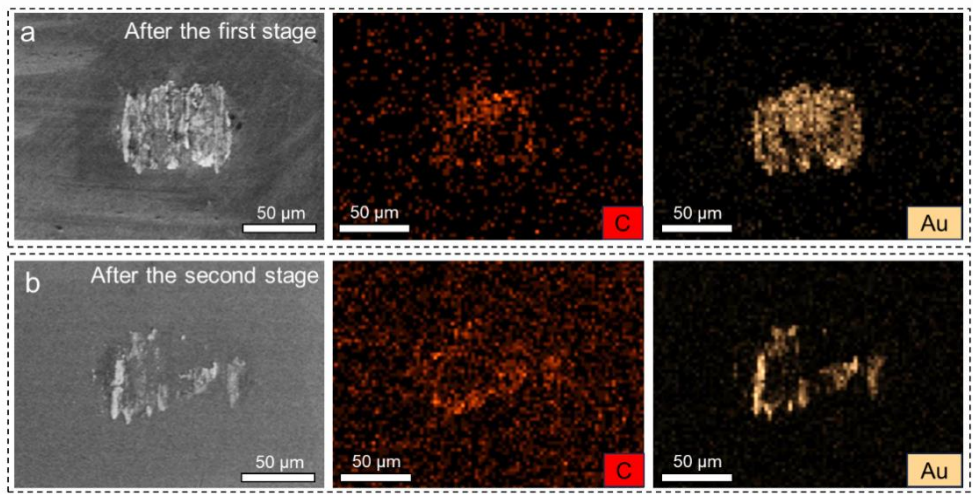

Supplementary Fig. 7| Metal replenishment behavior at the friction interface. a Metal-rich transfer films formed on the GCr15 counterpart ball after the first stage. b Metal-rich transfer films formed on the new GCr15 counterpart ball after the second stage.

**Supplementary Note 2: Frictional heat-induced metal migration.**

Herein, a static pressure experiment was conducted in which only stress was applied to the Cu/C film surface to observe whether the metal NPs migrate onto the film surface. As shown in Supplementary Fig. 8, a maximum Hertzian contact stress of 1.11 GPa was applied to the contact area. EDS analysis of the contact area reveals no significant Cu element aggregation (Supplementary Fig. 8c, d), and further XPS analysis also indicts that there is no significant increase in the Cu content on the contact area (Supplementary Fig. 8f). Nor is there any significant change in the carbon structure (Supplementary Fig. 8g).

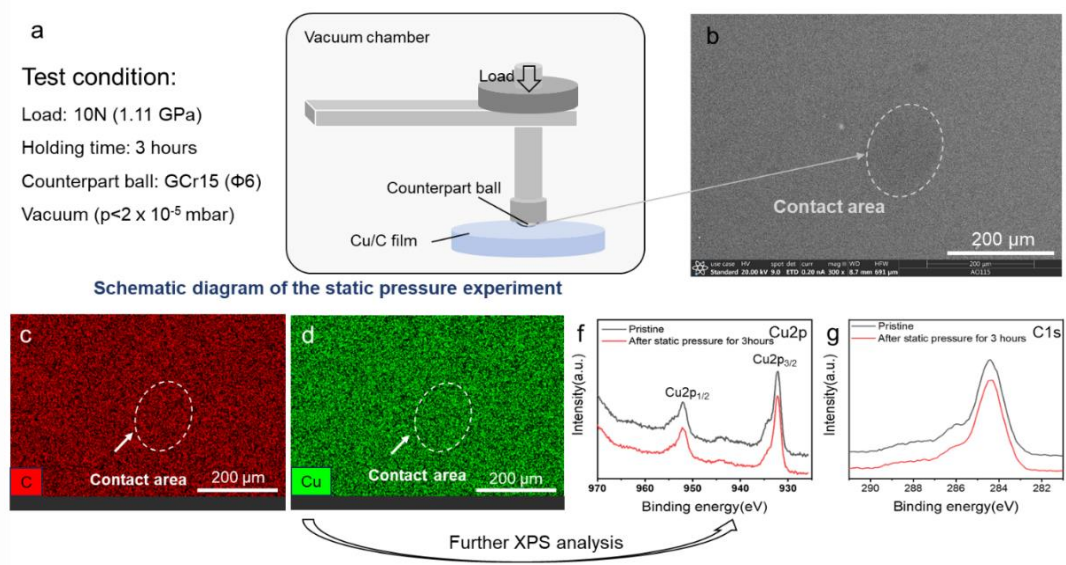

Supplementary Fig. 8| Static pressure experiment. a Schematic diagram and test conditions of the static pressure experiment. b FESEM image of the Cu/C film surface. c,d Element distribution mappings of the Cu and the C element, respectively. f XPS of Cu 2p before and after the static pressure experiment. g XPS of C 1s before and after static pressure experiment.

The effect of temperature on the migration behavior of Cu NPs in Cu/C films was investigated by annealing in vacuum (holding temperature for 3 h, pressure $<2×10^{-5}$ mbar). Supplementary Fig. 9a-d show the FESEM images of surface morphology of Cu/C films after annealing in different temperature. There are nano-pores present on the surface of the original film, which are inherent defects formed during the deposition. After annealing at 100 °C, the pores on the surface become larger. As the annealing temperature is increased to 200 °C, large white NPs are observed on the film surface, which also became denser without the large pores. At 300 °C, more white NPs are observable on the film surface, with the film structure remained dense. Elemental point scanning analysis of the white NPs are mainly composed of Cu elements. Further *in situ* XPS results indicate that the Cu/C ratios on the film surface gradually increase with rising temperature, which consistent with the trend of Cu NPs precipitation on the film surface after annealing at different temperature.

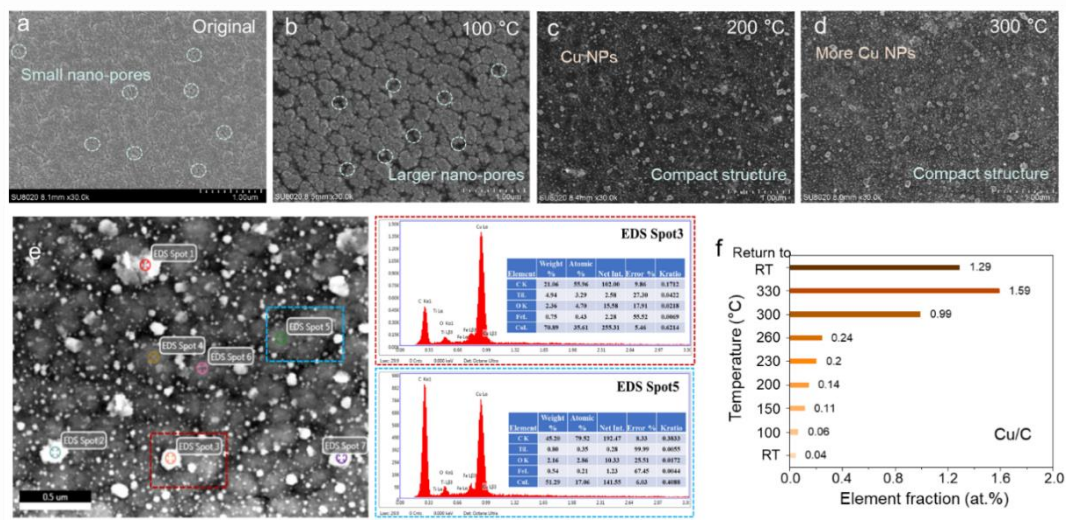

Supplementary Fig. 9| Heat-induce metal migration toward the film surface. a-d FESEM images of the surface of Cu/C films after annealing at different temperature in vacuum. e Point scanning of the surface of the Cu/C film after annealing at 300 °C in vacuum. f *In situ* XPS analysis of the Cu/C ratio evolution on the film surface during heating.

Real-time observation of the migration process of Cu NPs in amorphous carbon (a-C) matrices during heating was obtained by *in situ* TEM. There is no listing of the cross-sectional TEM images of Cu/C films due to no obvious structural changes of the films before 180 °C. Atomic thermal vibration intensifies at around 200 °C, Cu atoms dispersed in the a-C matrices come close to aggregation and start to nucleate or aggregate towards other Cu NPs and then grow up to make the system more stable. When the temperature increases to 300 °C, the size of Cu NPs is stable and spherical.

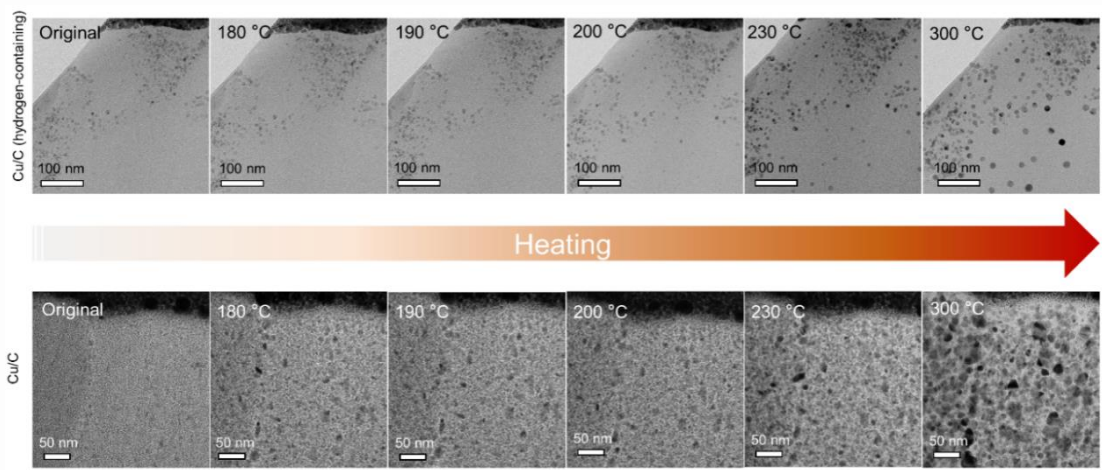

Supplementary Fig. 10| *In situ* TEM images of the migration process of Cu NPs in the a-C matrices during heating.

Supplementary Fig. 11 shows the original microstructures of Cu/C (hydrogen-free) film and Cu/C (hydrogen-containing) film, respectively. Cu nanocrystals (2~5 nm) are uniformly dispersed as metallic Cu in both films without oxidation. Cu NPs near the surface are partially oxidized to form $Cu_2O$ from Supplementary Fig. 23, which indicates that although the Cu element on the surface is partially oxidized during the storage process, the internal Cu element has not been oxidized due to the protective effect of the a-C matrix. Therefore, the effect of $Cu_2O$ has not been considered in this paper. Less Cu is incorporated due to hydrogen passivation of carbon dangling bonds in Cu/C (hydrogen-containing) films. While there are significant amounts of Cu NPs in Cu/C(hydrogen-free) films.

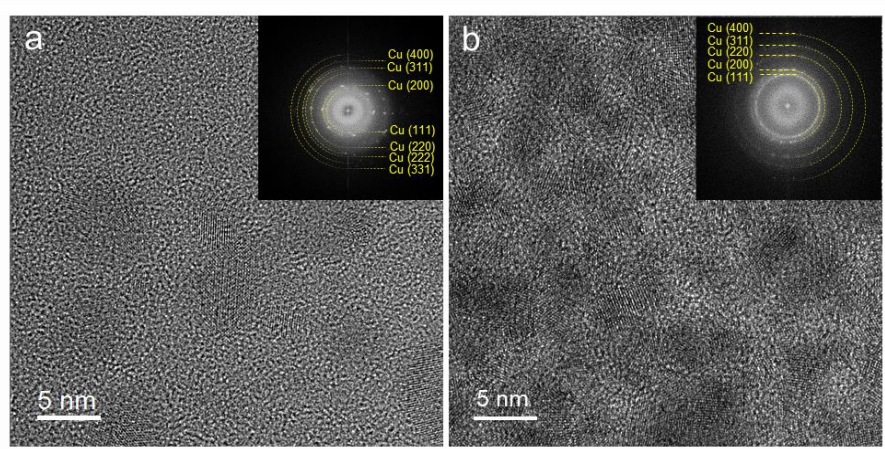

Supplementary Fig. 11| In situ TEM images of the original microstructures of Cu/C films. a The Cu/C (hydrogen-containing) film (previously prepared Cu0.75 A/a-C:H film[1]). b The Cu/C (hydrogen-free) film.

**Supplementary Note 3: The studies of metal migration mechanism.**

Supplementary Fig. 12 illustrates the physical phase transition process of Cu NPs during *in situ* heating. Consistent with Supplementary Fig. 10, newly grown Cu NPs were observed at 300 °C. When the temperature increased to 400 °C, Cu NPs from near the upper surface of the Cu/C film start to "disappear" gradually.

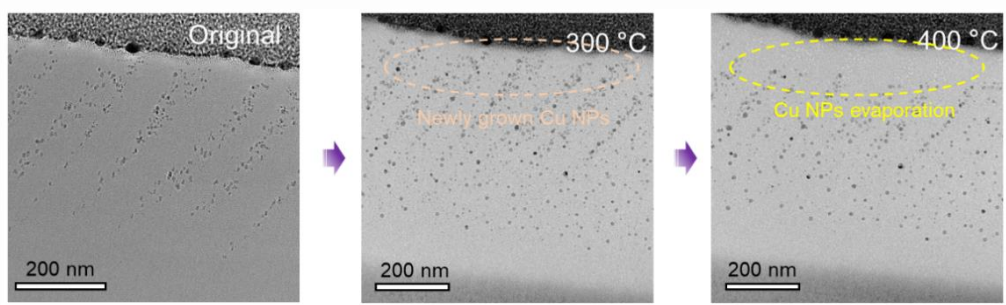

Supplementary Fig. 12| *In situ* TEM images of the evaporation of Cu NPs in the a-C matrices during heating.

Supplementary Fig. 13 presents the Cu 2p of the Cu/C film powders before and after the DSC-TG test. Before the DSC-TG test, obvious Cu signal is observed. However, it is unable to detect Cu signal after the DSC-TG test. The results demonstrate the loss of Cu during the DSC-TG test. Combining with Supplementary Fig. 12, these results indicate that Cu NPs evaporated due to excessively high temperatures, leading to the meltdown of their crystal structure.

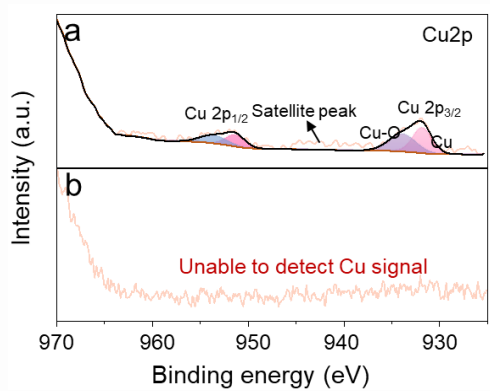

Supplementary Fig. 13| Cu 2p of the Cu/C film powders before and after DSC-TG test

Supplementary Fig. 14 illustrates the location of the Cu NP in the Molecular dynamics (MD) simulation. The displacement of the Cu NP along the z-axis over time shows the movement of the Cu NP toward the surface.

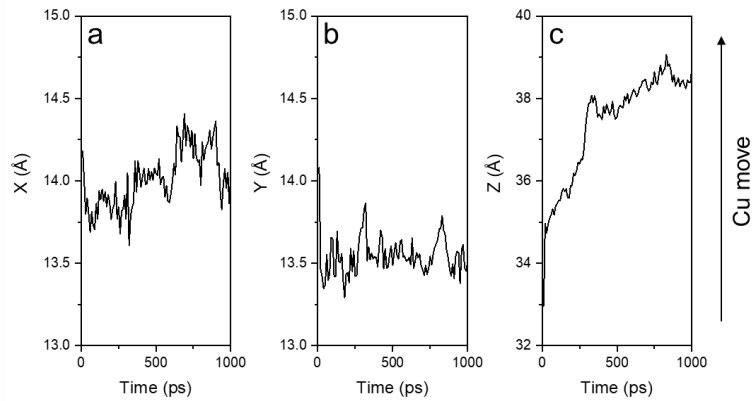

Supplementary Fig. 14| The location of the Cu NP in the MD simulation of the metal migration.

Supplementary Fig. 15 shows the gradient temperature distribution with along the friction interface[2]. The heat generated in the friction process increases the surface temperature of the films and forms a semi-spherical isothermal surface near the contact point. The friction heat is generated in the outermost deformation zone, so the surface temperature is the highest in θs. Under heat conduction, the deformation zone has a temperature gradient, and the matrix temperature outside the deformation zone changes gently θv.

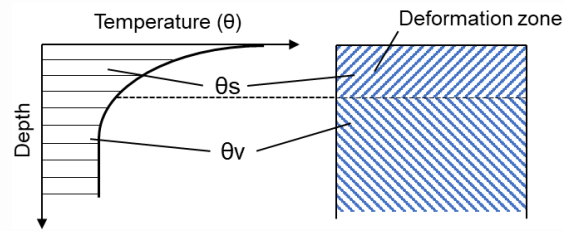

Supplementary Fig. 15| The isotherm inside the surface.

After annealing at 300 °C, numerous Cu nanoparticles precipitated on the film surface (Supplementary Fig. 9). Interestingly, no detrimental effect on the hardness or elastic modulus of Cu/C films was observed after the annealing process.

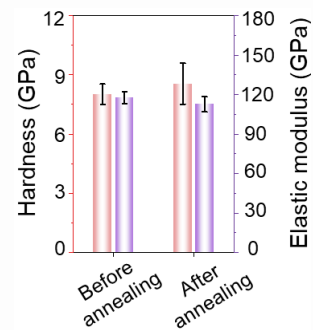

Supplementary Fig. 16| Mechanical properties of the Cu/C film before and after annealing in vacuum at 300 °C.

**Supplementary Note 4: The mechanism of ordered carbon formation catalyzed by metals.**

The friction interface of the Cu/C films in the friction stabilization stage are shown in Supplementary Fig. 17, The Cu/C film does not wear out and forms dense transfer films on the GCr15 counterpart ball, while C film fails quickly. Further elemental analysis of the wear scar of the Cu/C film was conducted. The wear scars primarily consist of Cu and C elements. Moreover, the Cu content during the stable period is significantly higher than that during the run-in period. The above results indicate that metals transfer onto the counterpart balls during friction, forming metal-rich transfer films, enabling the Cu/C film exhibits excellent tribological properties in vacuum.

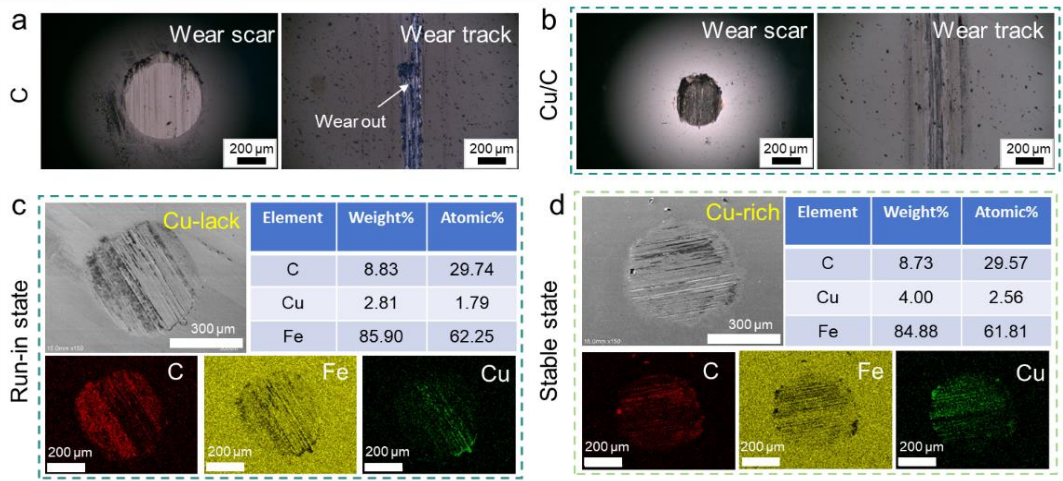

Supplementary Fig. 17| The friction interface of Cu/C films in the friction stabilization stage. a, b optical morphology of wear scars and wear tracks of the C film and the Cu/C film. c, d Element distribution mapping of the wear scars of the Cu/C film during the run-in (high $\mu$) and stable (low $\mu$) periods

The Raman signal of a-C is used to reflect the stability of the interfacial lubrication structure. The spectra of the Cu/C film were fitted with D and G peaks, corresponding to the respiration mode of the $sp^2$C atoms in the ring structure and the stretching of the $sp^2$C atoms in the ring and chain structures, respectively[3]. Structural Changes in the Cu/ C film can be qualitatively represented by the ratio of the areas of the two peaks ($I_D/I_G$). As shown in Supplementary Fig. 18, the $I_D/I_G$ ratio fluctuates drastically during the run-in period, indicating severe structural evolution and an unstable friction interface. Conversely, the ratio stabilizes during the steady-state friction period, signifying a robust interfacial structure. This stability is attributed to the formation of Cu-rich transfer films (Supplementary Fig. 17), where Cu NPs migrate to the friction interface, forming a stable lubrication structure.

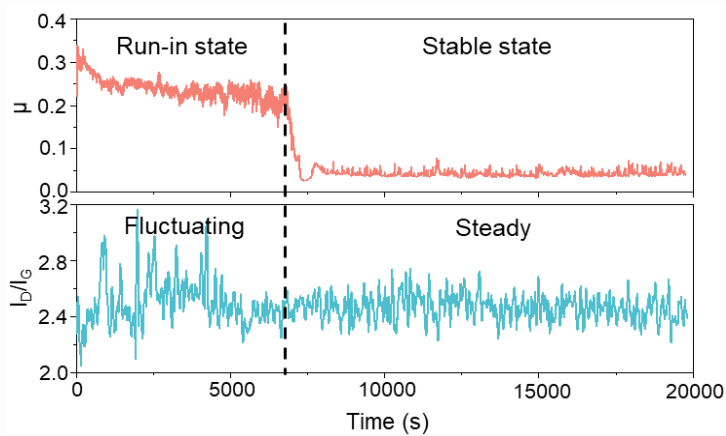

Supplementary Fig. 18| Correlations between μ and $I_D/I_G$ of the Cu/C film detected by *in situ* Raman during friction in vacuum.

Supplementary Fig. 19 shows the differential charge density (DCD) maps of the Cu-C configuration before the catalytic reaction. The pronounced charge redistribution indicates net electron transfer from Cu to C, consistent with their electronegativity difference.

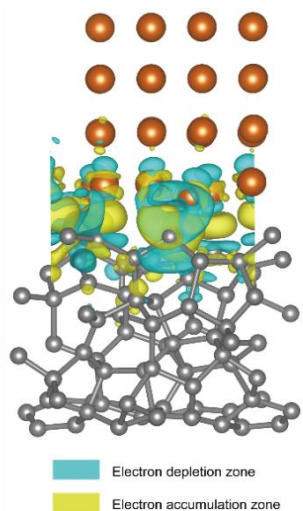

Supplementary Fig. 19| DCD maps of the Cu-C configuration before the catalytic reaction.

Supplementary Fig. 20 shows the -COHP of the Cu-C bond. By convention, positive -COHP values indicate bonding interactions, whereas negative values signify antibonding interactions. The integrated COHP (ICOHP), evaluated up to the Fermi level, provides a quantitative measure of bond strength. Therefore, a negative -COHP near the Fermi level of the Cu-C bond indicates the formation of Cu-C anti-bonds.

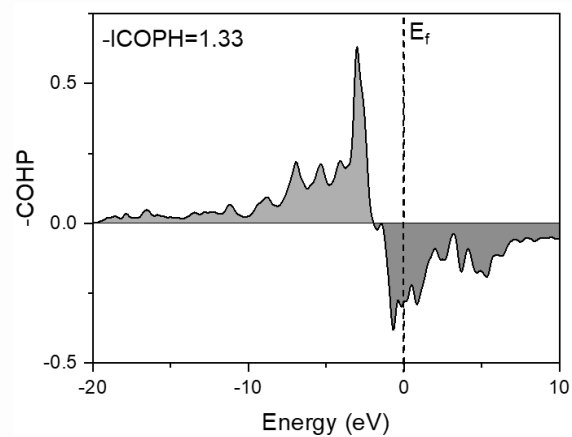

Supplementary Fig. 20| -COHP of the Cu-C bond.

To interrogate the electronic driving forces, we computed the work functions of Cu and a-C surfaces. The work function of Cu (4.62 eV) is substantially lower than that of a-C (6.05 eV), implying a greater propensity for electron donation from Cu.

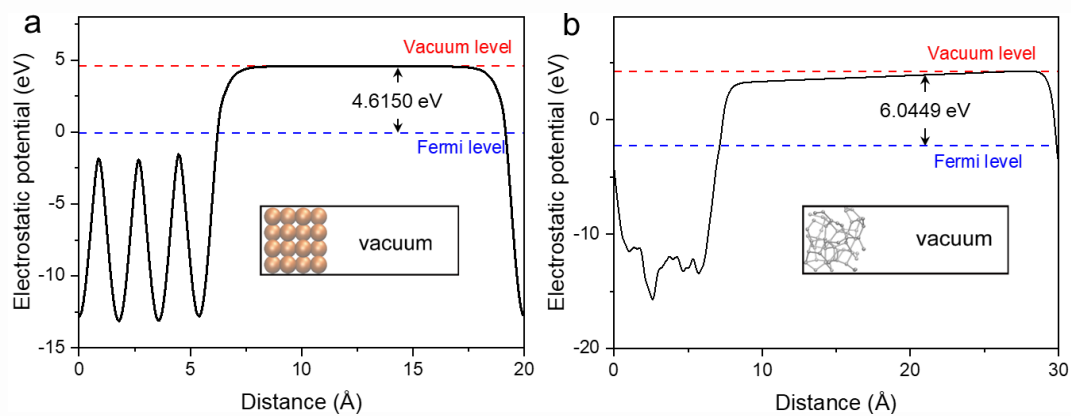

Supplementary Fig. 21| Work function calculations for the individual components. a Work function of the Cu system. b Work function of the a-C system.

**Supplementary Note 6: The preparation and information of the Cu/C films.**

The deposition process and parameters of Cu/C films are presented in Supplementary Fig. 22a. Cu/C films are prepared by closed-field non-equilibrium magnetron preparation technique, containing two graphite targets (99.99% purity), one Ti target (99.99% purity), and one Cu target (99.99% purity). The Cu/C films and C films were prepared respectively by modulating the Cu sputtering current (0A and 0.5A). Prior to deposition, the contaminations on the surface of the substrates were etched away by argon plasma pulsed bias of -400 V for 40 mins. Then, the Ti and TiC interlayers were deposited to enhance the adhesion between the Cu/C films and the substrates. The rotation speed is 5 r/min. Supplementary Fig. 22b, c presents FESEM images of the surface and cross-sectional morphology of C films and Cu/C films. There are nano-pores present on the surface, which are inherent defects formed during the deposition. The films possess high densification. The film thickness and the Cu content in the Cu/C film are labeled in Supplementary Fig. 22b, c.

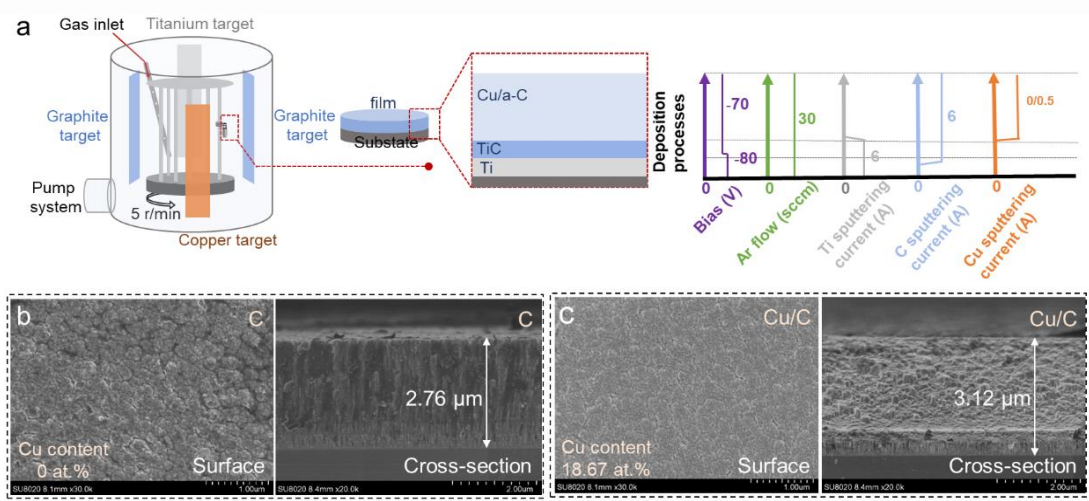

Supplementary Fig. 22| Deposition Process and information about Cu/C films. a Deposition Process and parameters. b, c FESEM images of surface and cross morphology of C film and Cu/C films.

The hardness and elastic modulus of the C films are 10.36 GPa and 140.93 GPa. The hardness and elastic modulus of the Cu/C films decrease due to incorporation of soft Cu, which are 8.01 GPa and 117.75 GPa. Supplementary Fig. 23 shows the scratch morphology of the C film and Cu/C film, the position marked by the red line is the initial peeling point, and the force corresponding to this point is the criterion of the adhesion between films and substrates[4]. The results indicate that the Cu/C film possess a great adhesive strength of the substrates.

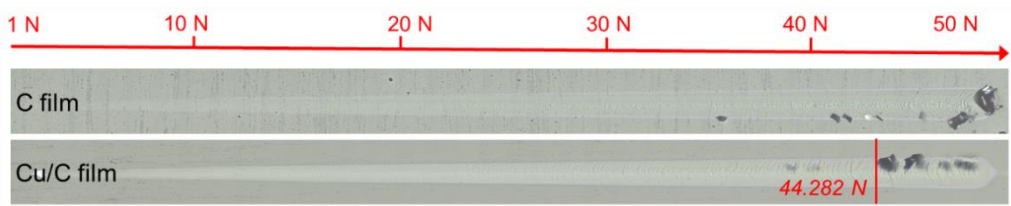

Supplementary Fig. 23| Scratch morphology of Cu/C films.

The Raman spectra of Cu/C films is shown in Supplementary Fig. 24a. XRD patterns of Cu/C films are depicted in Supplementary Fig. 24b, the diffraction peaks of Cu are wide and short, which indicates that these Cu exists in a-C matrices as nanocrystals with small sizes. The XPS of C1s, O1s, Cu 2p, and Cu LMM of Cu/C film are shown in Supplementary Fig. 24c-f. The Cu in the Cu/C films exists in the form of Cu and $Cu_2O$, and there is no chemical bonding between Cu-C. Combined with Supplementary Fig. 11, the localized oxidation exists due to the Cu NPs on the surface which are placed in the atmospheric environment due to higher chemical activity of Cu NPs on the surface.

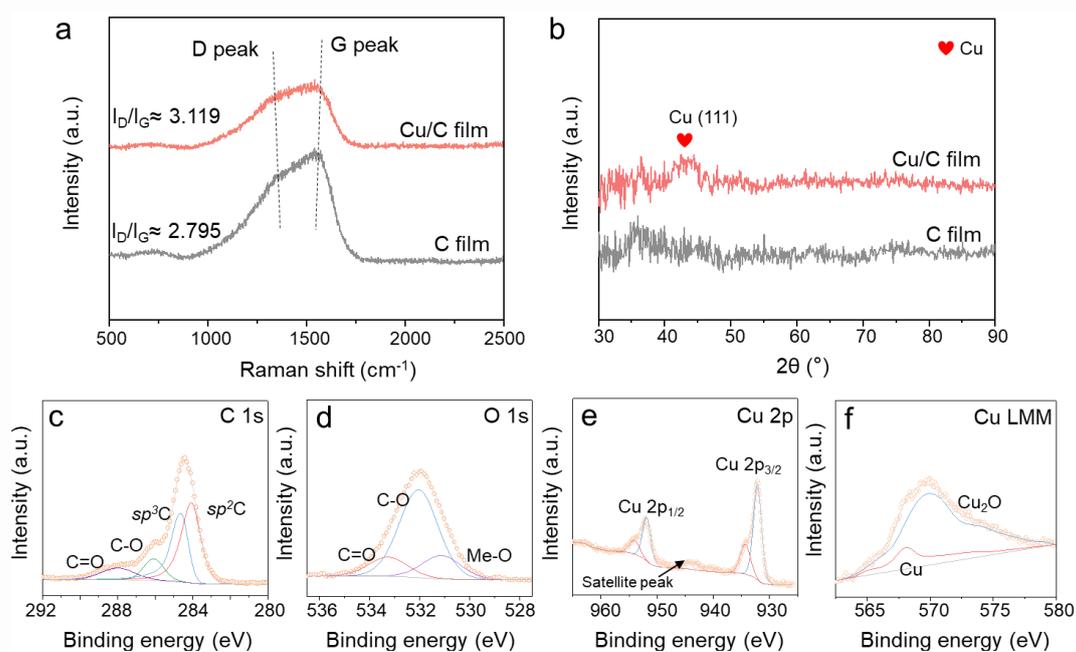

Supplementary Fig. 24| The original structure of Cu/C films. a Raman spectrum. b XRD patterns. c-f XPS of the Cu/C film.

**Supplementary Note 7: Development of the NEP model and set-up for MD simulations.**

The trained Neuroevolution potential (NEP) model for Cu-C overall performs well in predictions of energy, force and virial, with the test errors of RMSEe = 15.658 meV/atom, RMSEf = 0.311 eV/Å, and RMSEv = 57.159 meV/atom, respectively.

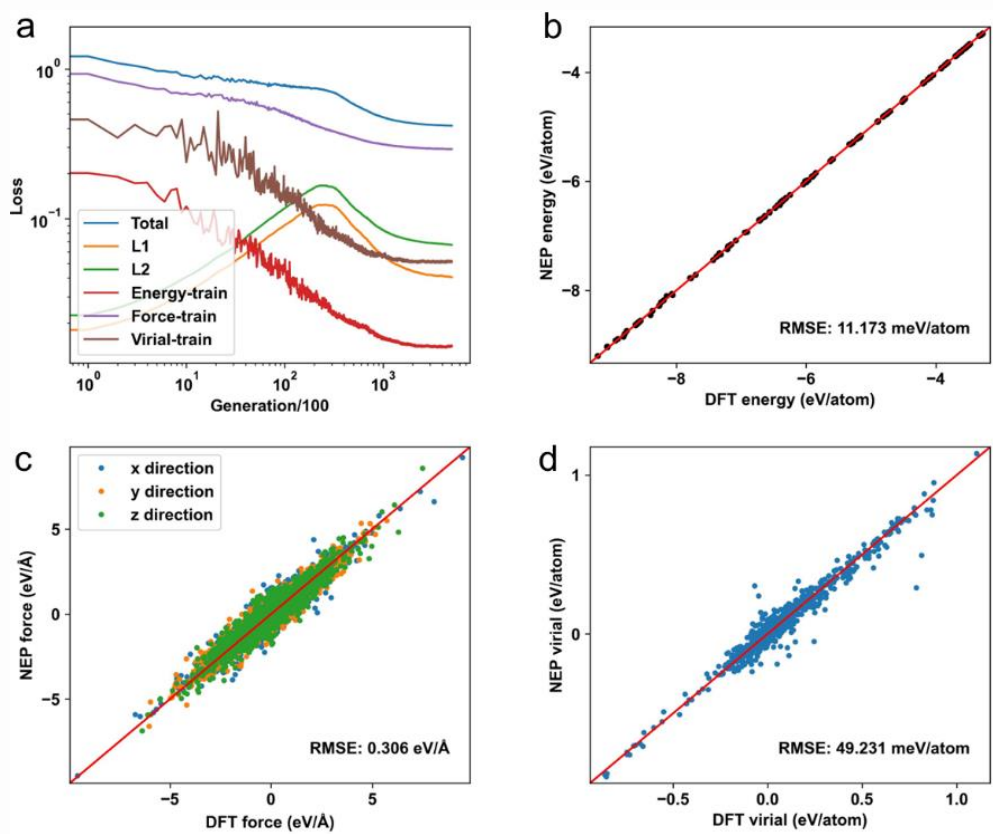

Supplementary Fig. 25| Trained NEP model. a Evolution of loss functions of energy (eV/atom), force (eV/Å) and virial (eV/atom) RMSEs during training generation for training set. b Parities of energy. c Parities of force. d Virial calculated from the NEP model as compared to the DFT-PBE reference data for the test set.

Supplementary Table. 1| Hyperparameters for NEP training.

| | |
|---|---|
| version | 4 |
| type | 2 Cu C |
| zbl | 2.0 |
| cutoff | 6 4 |
| n_max | 8 8 |
| basis_size | 12 12 |
| l_max | 4 2 |
| neuron | 50 |
| lambda_1 | 0.05 |
| lambda_2 | 0.05 |
| lambda_e | 1.0 |
| lambda_f | 1.0 |
| lambda_v | 0.1 |
| batch | 1000 |
| population | 50 |
| generation | 500000 |

We construct a a-C model by using different melt-quench temperature method for migration simulation. Frist, a diamond 8 × 8 × 20 supercell (containing 10,240 atoms) is heated to 9,000 K under NVT ensemble conditions for 400 ps to form liquid carbon sample. The liquid carbon sample is cooled down to lower temperature at 5,000 K and equilibrated for 200 ps. We then perform a rapid melt-quench process from 5,000 K to 1,000 K with a rate α = 1,012 K/s to form disordered a-C sample. Finally, the system is quenched from 1,000 K to 300 K over 200 ps and equilibrated at 300 K for 500 ps to ensure the complete elimination of metastable configurations and residual internal stresses. The radial distribution function (RDF) curve (see Supplementary Fig. 26) exhibits distinct peaks at interatomic distances of $r_1$ = 1.42 Å and $r_2$ = 2.46 Å, which are characteristic of a short-range ordered structure and further indicate the presence of $sp^2$-carbon domains within the a-C sample.

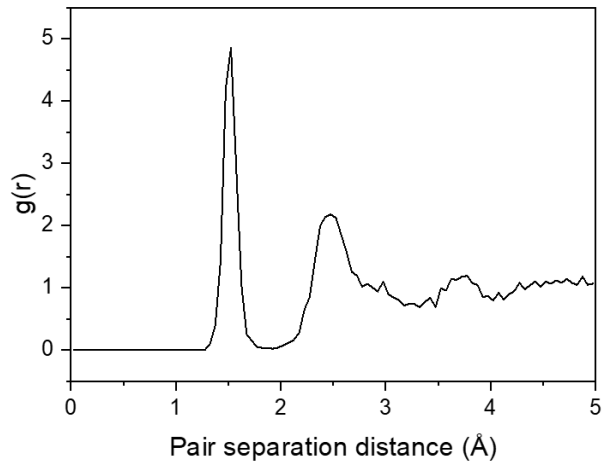

Supplementary Fig. 26| The RDF curve of the a-C sample.